\documentclass[journal]{IEEEtran}
\usepackage{draftwatermark}
\SetWatermarkText{Accepted}
\SetWatermarkScale{0.8}
\SetWatermarkColor[gray]{0.85}
\usepackage{graphicx}
\usepackage{multirow}

\usepackage{amsfonts}
\usepackage[numbers,sort&compress,square]{natbib}
\usepackage{siunitx}
\usepackage{optidef}
\ifCLASSINFOpdf
\else
\fi
\usepackage{lineno}
\usepackage{epstopdf}
\usepackage{nomencl}
\makenomenclature
\usepackage{algorithm}
\usepackage{algorithmic}
\usepackage{amstext}
\usepackage{amsmath}
\usepackage{amsfonts}
\usepackage{color}
\usepackage{epstopdf}
\usepackage{array}
\usepackage[normalem]{ulem}
\usepackage{nomencl}
\usepackage{mathrsfs}

\usepackage{subcaption}
\usepackage{graphicx}  
\usepackage{soul}
\usepackage{physics}

\begin{document}

\title{Coordination of OLTC and Smart Inverters for Optimal Voltage Regulation of Unbalanced Distribution Networks}


\author{\IEEEauthorblockN{Changfu Li, \textit{Student Member, IEEE}, Vahid R. Disfani, \textit{Member, IEEE}, Hamed Valizadeh Haghi, \textit{Senior Member, IEEE} and Jan Kleissl}

\thanks{A preliminary version of this paper is accepted for the 2019 IEEE PES General Meeting,
Atlanta, GA, 2019 \cite{li2019optimal}. Paper no. 19PESGM0236.} 

\thanks{Changfu Li, Hamed Valizadeh Haghi and Jan Kleissl are with the Center for Energy Reserch and the Department of Mechanical and Aerospace Engineering at the University of California San Diego, La Jolla, CA 92093 USA. \{emails: chl447@ucsd.edu, valizadeh@ieee.org,
jkleissl@ucsd.edu\}}

\thanks{Vahid R. Disfani is with the Department of Electrical Engineering at the University of Tennessee at Chattanooga, Chattanooga, TN 37403 USA. \{email: vahid-disfani@utc.edu\}}
}

\maketitle
\IEEEpeerreviewmaketitle

\begin{abstract}
Photovoltaic (PV) smart inverters can improve the voltage profile of distribution networks. A multi-objective optimization framework for coordination of reactive power injection of smart inverters and tap operations of on-load tap changers (OLTCs) for multi-phase unbalanced distribution systems is proposed. The optimization objective is to minimize voltage deviations and the number of tap operations simultaneously. A novel linearization method is proposed to linearize power flow equations and to convexify the problem, which guarantees convergence of the optimization and less computation costs. The optimization is modeled and solved using mixed-integer linear programming (MILP). The proposed method is validated against conventional rule-based autonomous voltage regulation (AVR) on the highly-unbalanced modified IEEE 37 bus test system and a large California utility feeder. Simulation results show that the proposed method accurately estimates feeder voltage, significantly reduces voltage deviations, mitigates over-voltage problems, and reduces voltage unbalance while eliminating unnecessary tap operations. The robustness of the method is validated against various levels of forecast error. The computational efficiency and scalability of the  proposed approach are also demonstrated through the simulations on the large utility feeder.
\end{abstract}

\section{Introduction}
Penetration of variable distributed generation connecting into distribution systems has  increased significantly in recent years. In California, specifically solar photovoltaics (PV) has shown such growth predominantly. While PV brings economic and environmental benefits, it presents voltage regulation challenges in distribution systems due to the variability in the solar resource \cite{Yan2012,Liu2012,li2019rl,cristian19}. 

Conventionally, utility devices such as on-load tap changers (OLTCs), shunt capacitors (ShCs), and shunt reactors regulate the voltage within operation limits. These devices are usually limited in number of switches and slow in response time, and hence less effective in regulating feeder voltage during periods of minute-by-minute PV variability. In contrast, PV smart inverters (SIs) provide an alternative method for fast-response voltage regulation by modulating real and/or reactive power of PV systems \cite{pecenak2017smart}. 
Moreover, all these devices typically operate autonomously based on pre-defined rules or curves to regulate voltage. These autonomous control schemes are based on local measurements requiring no communications. The lack of coordination between these devices leads to sub-optimal system performance. 

Numerous research works in the literature have studied coordination of SIs for voltage regulation \cite{farivar2012optimal, vsulc2014optimal,robbins2016optimal,dall2014decentralized, DallAnese2014,  guggilam2016scalable}. 
In \cite{farivar2012optimal}, optimal power flow (OPF) problems is formulated considering feeder-wide constraints and then solved to determine the optimal real and/or reactive power dispatch of SIs for minimizing losses while eliminating voltage violations. Through convex relaxation, the OPF is formulated as a second order cone program, and reactive power of SIs is optimized to reduce line losses and energy consumption. The alternating direction method of multipliers (ADMM) is used to solve the OPF and find the optimal SI reactive power for reducing losses in \cite{vsulc2014optimal} and for voltage regulation in \cite{robbins2016optimal}.  ADMM-based algorithms are also employed in \cite{dall2014decentralized} to determine optimal SI real and reactive power set points for voltage regulation. In \cite{DallAnese2014}, semi-definite programming relaxation is leveraged for optimal dispatch of SI real and reactive power. A linear approximation of the power flow equations is used in \cite{guggilam2016scalable} to optimizate SI real and reactive powers efficiently. In these works, however, cooperation amongst SIs is studied without taking other voltage regulation devices including OLTCs into account. 
Uncoordinated operation of OLTCs and SIs may cause unintended OLTC tap operations \cite{kraiczy2017parallel} leading to higher OLTC wear and tear and less effective voltage regulation. 

Various studies have improved coordination between SIs and other voltage regulation devices for better voltage regulation performance. Some works improve the basic autonomous rule based methods \cite{ku2018coordination, maruf2019impact, singh2019coordinated}. Reference \cite{ku2018coordination}  replaces the local OLTC bus voltage with feeder end measurement for the control of tap switching to improve the visibility of downstream voltage. SIs dynamically  adjusts their power factor per autonomous curves. OLTC voltage set points are dynamically adjusted based on voltage estimates from sensitivity matrix to accommodate SI outputs in \cite{maruf2019impact}. OLTC and SIs are coordinated by iteratively updating their settings to achieve target local voltage at the SI \cite{singh2019coordinated}.

The improved rule based methods are relatively simple to implement and can accomplish partial coordination between devices. However, optimization based approaches can achieve optimal coordination to ensure optimal voltage regulation performances for distribution feeders with complicated voltage profiles caused by fluctuating distributed energy resources \cite{Christakou2013,christakou2017voltage,borghetti2010short,nguyen2018exact,othman2019coordinated}. In \cite{Christakou2013}, SI real and reactive power and OLTC tap positions are optimized simultaneously to minimize voltage deviations. A robust optimization model is developed in \cite{christakou2017voltage} to schedule real and reactive power of distributed generators (DGs) and OLTCs for production cost minimization while ensuring acceptable network voltage profiles. DGs and OLTCs are coordinated through optimization to minimize voltage deviations
and network losses in \cite{borghetti2010short}. The optimization in \cite{nguyen2018exact} also coordinates SIs, OLTCs, and ShCs while meeting voltage operation limit constraints. Reactive power of SIs, ShCs and OLTC tap position are optimized to eliminate voltage violations in  \cite{othman2019coordinated}.

Since non-linear AC power flow constraints render optimization problems non-convex and computationally intensive for large distribution networks, 
different linearization techniques have been applied in the literature to address this concern. 
In \cite{Christakou2013}, an analytical approach is proposed to calculate sensitivity coefficients of node voltages to approximate voltage change as a function of SI real power, reactive power, and OLTC tap positions. Since \cite{Christakou2013} assumes that OLTCs are located at the substation, the method is not directly applicable to distribution feeder with OLTC in the middle of the feeder \cite{Christakou2013,christakou2017voltage}. 
In \cite{borghetti2010short}, the sensitivity coefficients for OLTC tap position are determined by summing real and reactive power coefficients at all buses due to each OLTC variation. The computational burden of the method could increase substantially due to exponentially growing tap position combinations with more OLTCs. Linearized power flow equations are exploited for computational efficiency in \cite{guggilam2016scalable}. The solution, however, does not coordinate SIs and OLTCs and is not validated on multi-phase and unbalanced distribution feeders. Furthermore, the voltage estimate from the linear approximation differs substantially from the actual system voltage. 

SIs, OLTCs, and ShCs are coordinated in \cite{nguyen2018exact} and the method is tested on unbalanced feeders. The optimization problem is non-linear and non-convex without relaxation of AC power flow constraints. While computational strategies are introduced to reduce computation cost, this non-linear and non-convex formulation could lead to either local solutions or convergence issues \cite{nguyen2018exact,capitanescu2013experiments,capitanescu2016critical}. In \cite{othman2019coordinated}, the big bang-big crunch optimization is used to improve convergence speed. However, the computation time can still be up to 40$~$s for the small IEEE 33-bus feeder with only 3 DGs, 1 OLTC and 2 ShCs, which makes the optimization technique not applicable for distribution systems with hundreds and thousands of nodes.

In the authors' prior work \cite{Li2018}, coordination of multiple OLTCs for voltage regulation is studied. Voltage violations are mitigated and the method is proven to be computationally efficient. However, it does not address coordination between OLTCs and SIs, which can lead to the higher OLTC wear and less effective voltage regulation as discussed in \cite{kraiczy2017parallel}. 

As a follow-on work, this paper proposes an optimization-based voltage control strategy to coordinate OLTCs and SIs in operable time scales. It proposes 
a new linearization method to linearize power flow equations and to convexify the problem, which guarantees convergence of the optimization and less computation costs. The optimization  is  modeled  and  solved  using  mixed-integer  linear programming (MILP). Since the linearization technique convexifies the optimization problem, convergence are guaranteed in contrary to \cite{nguyen2018exact}. By relaxing the non-linear AC power flow constraints, fast solutions can be achieved in a regular PC, unlike \cite{othman2019coordinated}. Also, the proposed method addresses OLTCs which are located within the feeder--not necessarily on substation--which differentiates this paper from \cite{Christakou2013,christakou2017voltage}. The proposed formulation is scalable and can be easily applied to distribution networks with any number of OLTCs in contrast to \cite{borghetti2010short}. A sensitivity study shows that voltage estimation is more accurate with a maximum error of 0.009 p.u.  compared to around 0.1 p.u. in \cite{guggilam2016scalable}. The method is also applicable to unbalanced feeders. It is demonstrated on the highly-unbalanced modified IEEE 37 bus test network. And it is robust against forecast errors as demonstrated by simulations. The scalability of the method is also tested on a real California utility feeder with 2844 nodes.

In summary, the contributions of this paper are as follows:
\begin{enumerate}
    \item it addresses coordination of OLTCs and SIs for optimal voltage regulation,
    \item a novel linearization technique is proposed to convexify the optimization problem for higher computational efficiency,
    \item it guarantees convergence and leads to more accurate voltage estimates,
    \item it is robust against forecast errors,
    \item it is scalable to handle multiple OLTCs and SIs coordination regardless of OLTC location.
\end{enumerate}

The rest of the paper is organized as follows. Section \ref{sec:LMTOVI} discusses the linearization technique for modeling the OLTC tap change and SI reactive power on distribution feeder voltage. Section \ref{sec:opt} explains the formulation of the optimization. Section \ref{sec:casestudy} provides details of the test feeder models, voltage regulation methods, and simulation scenarios. Section \ref{sec:result} presents simulation results followed by conclusions in section \ref{sec:conc}. 

\section{Model Linearization for Voltage Regulation}
\label{sec:LMTOVI}

The goal of the optimization formulated in section~\ref{sec:opt} is to coordinate OLTCs and SIs for voltage regulation. In this section, we introduce the linearized model to represent the relation between voltage and controllable parameters.

\subsection{Linearization of Feeder Nodal Equation} 
Consider a distribution feeder with $N$ nodes contained in the set $\mathcal{N}$. Its feeder nodal voltage equation can be written as: 
\begin{align}
\textbf{V}=\textbf{Z}\textbf{I}, 
\label{eq:V-I}
\end{align}
where $\textbf{V}$ and $\textbf{I}$ are $N\times1$ complex vector for voltages and net node current injections at all nodes ($\textbf{V},\textbf{I}\in C^N$), and $\textbf{Z}$ is the $N\times N$ feeder impedance matrix ($\textbf{Z}\in C^{N\times N}$). 
A linear approximation of the perturbations in node voltage resulting from changes in impedance and current ($\partial \textbf{V} / \partial (\textbf{ZI})$) leads to
\begin{align}
\Delta \textbf{V}=\Delta \textbf{Z}\cdot \textbf{I}_0+\textbf{Z}_0 \cdot \Delta \textbf{I},
\label{eq:VI_derivaive}
\end{align}
where the subscript $(0)$ represents unperturbed parameters.
$\Delta \textbf{Z}$ is a function of tap position changes of OLTCs and $\Delta \textbf{I}$ results from current injection changes of PVs and loads. $\Delta \textbf{V}$ is further derived by modeling the effects of OLTC tap changes on $\Delta \textbf{Z}$ and current source changes on $\Delta \textbf{I}$ which we cover in Section \ref{subsec:VoltEffectTO} and Section \ref{subsec:VoltEffectCurr}.

\subsection{Modeling OLTC Tap Operation Effects on Voltage}
\label{subsec:VoltEffectTO}

An OLTC regulates voltage via changing tap position $\tau$, which alters the ratio of the transformer secondary voltage with respect to the primary voltage (tap ratio $a$) and changes the impedance matrix $\textbf{Z}$. The tap ratio $a$ is a linear function of $\tau$,
\begin{align}
 a=1+\frac{\tau}{\tau_{\max}}(a_{\max}-1),  
 \label{eq:tapratio}
\end{align}
where $a_{\max}$ is the maximum tap ratio corresponding to the maximum tap position $\tau_{\max}$. 
OLTC tap operation effects on voltage can be determined by modeling its effects on $\textbf{Z}$, which is a function of tap ratio $a$.

In \eqref{eq:VI_derivaive}, $\Delta \textbf{Z}$ is needed to determine $\Delta \textbf{V}$. Considering an OLTC tap operation, which changes the tap ratio from $a_0$ to $a$, the corresponding impedance matrix change can be expressed as 
\begin{align}
\Delta \textbf{Z}=-\textbf{Y}_0^{-1}\cdot\Delta \textbf{Y}\cdot \textbf{Y}_0^{-1},
\label{eq:DeltaZ}
\end{align}
where $\textbf{Y}_0$ is the admittance matrix associated with the initial tap ratio $a_0$. $\Delta \textbf{Y}$ is the admittance change due to change of OLTC tap ratio from $a_0$ to $a$.

 Since $\Delta \textbf{Y}$ is the only unknown on the right-hand-side of \eqref{eq:DeltaZ}, $\Delta \textbf{Z}$ can be determined if the corresponding $\Delta \textbf{Y}$ can be modeled. Considering an OLTC connected between node $i$ of the primary side and node $j$ of the secondary side, only the elements associated with these two nodes are impacted by a OLTC tap change, i.e. only the following elements in $\Delta \textbf{Y}$ are non-zero:
\begin{align}
\Delta Y_{ii}=(a^2-a_0^2)/z_T, \label{dYii}\\
\Delta Y_{ji}=\Delta Y_{ij}=-(a-a_0)/z_T, \label{dYij} 
\end{align}
where $z_T$ is the equivalent impedance of the transformer on the winding connected to node $i$. The non-linearity in \eqref{dYii} can be removed by a Taylor series expansion for $a^2$ around $a_0$, yielding a linear expression,
\begin{align}
\Delta Y_{ii} = (2aa_0 - 2a_0^2)/z_T.
\label{dYii_lin}
\end{align}


 More details on the derivation of  \eqref{eq:DeltaZ} and the relationship between $\textbf{Y}$ and $a$ can be found in \cite{Li2018}. The derivation of \eqref{eq:DeltaZ} is based on the assumption of fixed current loads in \cite{Li2018}, which does not accurately represent common loads \cite{pecenak2018inversion}. In this paper, we apply a more common fixed power model for loads. However, the expression of \eqref{eq:DeltaZ} still applies since both $\Delta \textbf{Z}$ and $\Delta \textbf{Y}$ in 
 \eqref{eq:DeltaZ} are direct results of OLTC tap changes and are only functions of tap positions.

\subsection{Modeling Voltage Impacts of Current Sources}
\label{subsec:VoltEffectCurr}
PVs and loads are the current sources. A change in their injected currents ($\Delta \textbf{I}$) affects the feeder voltage profile per \eqref{eq:VI_derivaive}. The power injections of PVs and loads need to be specified for modeling their current injections into the feeder.

The power injections and current injections are related by
\begin{align}
\textbf{S} = \textbf{P}+j\textbf{Q} = \textbf{V}\textbf{I}^*,
\label{node_apparent_power1}
\end{align}
where $\textbf{S},\textbf{P},\textbf{Q} \in C^N$ are the vectors of complex power, real power, and reactive power injections at all nodes. $j:=\sqrt{-1}$. $\textbf{V}$ is the voltage vector and $\textbf{I}^*$ is the conjugate of the net current vector. Expressing the parameters as the initial value plus a perturbation, $\textbf{V} = \textbf{V}_0+\Delta \textbf{V}$ and $\textbf{I} = \textbf{I}_0+\Delta \textbf{I}$, \eqref{node_apparent_power1} can be rewritten as,
\begin{align}
    \textbf{S} = (\textbf{V}_0+\Delta \textbf{V})(\textbf{I}_0+\Delta \textbf{I})^*.
    \label{node_apparent_power2}
\end{align}
Eq.~\eqref{node_apparent_power2} sets up the relation between $\Delta \textbf{I}$ and the power injections of PVs and loads as needed for Eq.~\eqref{eq:VI_derivaive}.
 
Substituting the real and imagery parts of $\textbf{V}_0$, $\Delta \textbf{V}$, $\textbf{I}_0$ and $\Delta \textbf{I}$ into \eqref{node_apparent_power2} yields the real and reactive power injection as,   
\begin{align}
\textbf{P} = (\textbf{V}_{d0}+\Delta \textbf{V}_d)(\textbf{I}_{d0}+\Delta \textbf{I}_d)+(\textbf{V}_{q0}+\Delta \textbf{V}_q)(\textbf{I}_{q0}+\Delta \textbf{I}_q), 
\label{realpower2}
\\\textbf{Q} = (\textbf{V}_{q0}+\Delta \textbf{V}_q)(\textbf{I}_{d0}+\Delta \textbf{I}_d)-(\textbf{V}_{d0}+\Delta \textbf{V}_d)(\textbf{I}_{q0}+\Delta \textbf{I}_q). 
\label{reactivepower2}
\end{align}
where $\textbf{V}_0 = \textbf{V}_{d0}+j\textbf{V}_{q0}$, $\textbf{I}_0 = \textbf{I}_{d0}+j\textbf{I}_{q0}$, $\Delta \textbf{V} = \Delta \textbf{V}_d+j\Delta \textbf{V}_q$, and $\Delta \textbf{I} = \Delta \textbf{I}_d+j\Delta \textbf{I}_q$.

The unperturbed variables (subscript ``0'') are known. The  terms with $\Delta$ symbol are the unknowns to be solved in the optimization. Imposing constraints of $\textbf{P}$ and $\textbf{Q}$ directly would result in a non-convex optimization problem due to products of two unknown optimization parameters (e.g.  $\Delta \textbf{V}_d\Delta \textbf{I}_q$ in $\textbf{P}$). To convexify the problem, $\textbf{P}$ and $\textbf{Q}$ are linearized and the constraints are implemented using $\Delta \textbf{P}$ and $\Delta \textbf{Q}$ ($\textbf{P}= \textbf{P}_0+\Delta \textbf{P}+\textbf{P}_{\rm err}$, $\textbf{Q}= \textbf{Q}_0+\Delta \textbf{Q}+\textbf{Q}_{\rm err}$). 
Higher order non-convex square terms are dropped to yield
\begin{align}
\Delta \textbf{P} = \textbf{V}_{d0}\Delta \textbf{I}_d+\Delta \textbf{V}_d\textbf{I}_{d0}+\textbf{V}_{q0}\Delta \textbf{I}_q+\Delta \textbf{V}_q\textbf{I}_{q0},\\
\Delta \textbf{Q} = \textbf{V}_{q0}\Delta \textbf{I}_d+\Delta \textbf{V}_q\textbf{I}_{d0}-\textbf{V}_{d0}\Delta \textbf{I}_q-\Delta \textbf{V}_dI_{q0}. 
\end{align}
The higher-order non-convex terms constitute the real and reactive power errors $\textbf{P}_{\rm err} = \Delta \textbf{V}_d \Delta \textbf{I}_d+\Delta \textbf{V}_q\Delta \textbf{I}_q$ and $\textbf{Q}_{\rm err} = \Delta \textbf{V}_q\Delta \textbf{I}_d-\Delta \textbf{V}_d\Delta \textbf{I}_d$.

For load nodes, assuming a commonly used fixed power load model \cite{kersting2006distribution}, the load power injections remain unchanged despite the perturbations.  Therefore the power injection constraints for load nodes are
\begin{align}
    \Delta P_i=0,    \forall i\in \mathcal{N}^{l},
    \label{deltaP_load}
    \\\Delta Q_i=0, \forall i\in \mathcal{N}^{l},
    \label{deltaQ_load}
\end{align}
where $\mathcal{N}^l$ is the set of load nodes. 

\begin{figure} 
\centering
\includegraphics[width=0.4\textwidth]{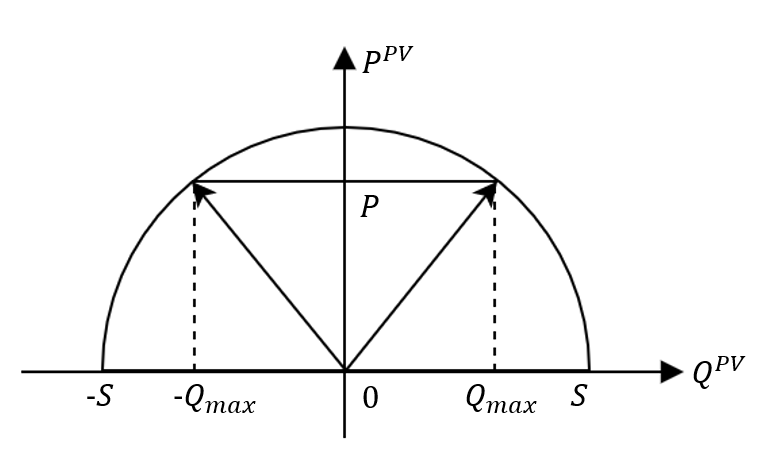}
\caption[]{PV smart inverter output curve. $P$ is the PV real power production, which is determined by the solar irradiance at each time step. Assuming no curtailment, given the SI rating $S$, the reactive power output can then vary between $-Q_{\rm max}$ and $Q_{\rm max}$.}
\label{fig:SI}
\end{figure}

For PV nodes, to maximize PV production real power curtailment is prohibited . The real power injections at the perturbed PV nodes then remain $P_{i0}, \forall i \in \mathcal{N}^{PV}$ and are determined by the available solar irradiance. The reactive power injections of the PV nodes are limited by the inverter rated power, $|Q_i|\leq Q_{i_{\rm max}}, \forall i \in \mathcal{N}^{PV}$ as shown in Fig.~\ref{fig:SI}. $Q_{i_{\rm max}} = \sqrt{S_i^2 - P_i^2}$ is the maximum available reactive power of the SI, where $S_i$ is the SI rated power, $\forall i \in \mathcal{N}^{PV}$. After the linearization, the constraints at the PV nodes become
\begin{align}
    \Delta P_i=0, \forall i \in \mathcal{N}^{PV},
    \label{deltaP_PV}\\
    |\Delta Q_i| \leq (Q_{i_{\rm max}} - Q_{i_0}), \forall i \in \mathcal{N}^{PV}.
    \label{deltaQ_PV1}
\end{align}
Assuming that the PVs operate at unity power factor before reactive power perturbations ($Q_{i_0} = 0$), the constraint in \eqref{deltaQ_PV1} becomes 
\begin{align}
-Q_{i_{\rm max}} \leq \Delta Q_i \leq Q_{i_{\rm max}},\forall i \in \mathcal{N}^{PV}.
\label{deltaQ_PV2}
\end{align}

\subsection{Linearization of voltage magnitudes}
The magnitude of the complex node voltages can be calculated based on their real and imaginary parts as
\begin{align}
|V|^2=V_{d}^2+ V_{q}^2,
\label{eq:vamag}
\end{align}
where $V = V_{d}+jV_{q}$ is the complex voltage of an arbitrary node of the feeder.

Linearizing the nodal voltage magnitude in \eqref{eq:vamag} around the initial point (i.e. $V_{0} = V_{d0}+jV_{q0}$) yields $|V_0|\Delta |V| = V_{d0}\Delta V_d+V_{q0}\Delta V_q$. Then the voltage magnitude of an arbitrary node can be calculated as
\begin{align}
|V|=|V_0|+\Delta|V|=|V_0|+|V_0|^{-1}(V_{d_0}\Delta V_d+V_{q_0}\Delta V_q).
\label{vamaglin}
\end{align}

This definition sets up an affine relation between the voltage magnitude of all nodes and the optimization parameters, which convexifies the optimization problem.

\subsection{Implementation and Forecasts}
\begin{figure} 
\centering
\includegraphics[width=0.46\textwidth]{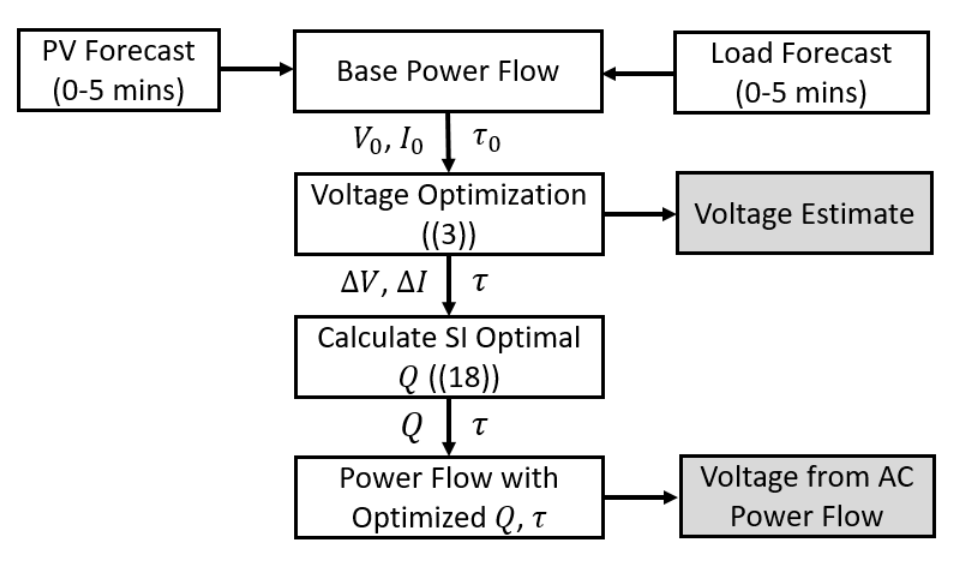}
\caption[Feeder Topologies]{Flowchart of the proposed voltage optimization. PV and load forecasts are used to obtain the linearization voltages and currents ($\textbf{V}_0$, $\textbf{I}_0$) for the next 5 minutes from an OpenDSS \cite{OpenDSS} base power flow simulations. Then, the voltage optimization per \eqref{optimization_problem} is formulated and solved by interfacing with the CVX \cite{cvx} and Gurobi solvers \cite{gurobi} using MATLAB, providing decision values for $\Delta \textbf{V}$, $\Delta \textbf{I}$ and $\tau$. The optimal SI reactive power of the SI per \eqref{reactivepower2} and the optimal tap positions $\tau$ are then input into another power flow. The voltage estimation accuracy using \eqref{vamaglin} is verified by comparing to the voltage results from the OpenDSS simulation.
}
\label{flowchart}
\end{figure}

Fig.~\ref{flowchart} presents the flowchart of the implementation of the proposed voltage optimization. For tap operation minimization, the optimization problem is defined over a 5-min time horizon. $\textbf{V}_0$ and $\textbf{I}_0$ over the next 5 minutes are needed for modeling the effects of OLTC tap position changes and SI reactive power on voltage and they are obtained from a base power flow run by OpenDSS \cite{OpenDSS} using solar and demand forecasts. Sky imagers provide forecasts of PV availability throughout the feeder at high spatio-temporal resolution \cite{yang2014solar}. Load profile is from measured data at the substation provided by the utility. Forecasts are generated from the original data by adding different levels of random noise. Details of forecast construction are elaborated in Section \ref{subsec:resforecast}. The optimal tap positions and SI reactive power set points are assumed to be delivered back to each OLTC/SI without communication delay.

\section{Feeder-Wide OLTC and SI Optimization}
\label{sec:opt}
In this section, we formulate the optimization for coordination of OLTCs and SIs. The goals are to mitigate voltage violations through minimizing voltage deviations and to reduce tap operations. The optimization objectives are formulated according to these two goals. 
\subsection{Optimization Model} 
\label{optimizationmodel}
The first objective function ($J_1$) is the sum of voltage deviations from $1$ p.u. on the feeder during the optimization horizon
\begin{align}
J_1=\sum_{i=1}^{N} \sum_{t\in T}(||V_i(t)|-1|),
\label{J1}
\end{align}
where $T$ is the set of time steps in the optimization horizon, and $|V_i(t)|$ denotes the voltage magnitude of node $i$ at time step $t$. Minimizing $J_1$ achieves a more homogeneous and steady voltage.

The second objective function ($J_2$) counts the number of 
tap operations as,
\begin{align}
J_2=\sum_{p\in P}\sum_{t\in T}|{\tau_{p,t+1}-\tau_{p,t}}|,
\label{J2}
\end{align}
where $P$ is the set of all OLTCs and $\tau_{p,t}$ denotes the tap position of OLTC $p$ at time step $t$. All tap changes over a defined time horizon $T$ are aggregated in $J_2$.

Combining the two objective functions, the optimization can be formulated as:

\begin{mini}
|s|
{}{J = w_1J_1 + w_2J_2}{}{}
\addConstraint{\eqref{eq:VI_derivaive}, \eqref{eq:tapratio}, \eqref{eq:DeltaZ}, \eqref{dYij}, \eqref{dYii_lin}}
\addConstraint{\eqref{deltaP_load}, \eqref{deltaQ_load}, \eqref{deltaP_PV}, \eqref{deltaQ_PV2}, \eqref{vamaglin}}
\addConstraint{\tau_{p,t}\in\mathbb{Z}}
\addConstraint{\tau_{p,\min}\le\tau_{p,t}\le\tau_{p,\max}}
\addConstraint{|\tau_{p,t}-\tau_{p,t-1}|\le \Delta TO_{p,\max}}
\end{mini}

The weighting factors $w_1$ and $w_2$ balance voltage regulation performance and total tap operations. Increasing $w_1$ will improve the voltage profile at the cost of more tap operations and vice versa. $w_1 = 1$ and $w_2 = 0.15$ are chosen in this paper through trial-and-error. This combination of parameters provide good performance on both test feeders. Details of weights factors selection are provided in Section \ref{sec:weightfactor}.

\subsection{Optimization Constraints}
\label{constraint}

As presented in Section \ref{optimizationmodel}, \eqref{eq:VI_derivaive} is included as an equality constraint to represent the linearized feeder nodal voltage equation. \eqref{eq:tapratio}\eqref{eq:DeltaZ}\eqref{dYij}\eqref{dYii_lin} are equality constraints for modeling the relationship between impedance matrix change ($\Delta Z$) and OLTC tap position ($\tau$).
\eqref{deltaP_load}\eqref{deltaQ_load} are the fixed power load model constraints for load real power and reactive power, respectively.

In this work, real power curtailment is prohibited to maximize PV production, \eqref{deltaP_PV} is the corresponding equality constraint. As discussed in Section \ref{subsec:VoltEffectCurr}, the maximum available SI reactive power injection is limited by the inverter rated capacity (Fig.\ref{fig:SI}). \eqref{deltaQ_PV2} reflects the constraint on reactive power injection for each SI. The linearized voltage magnitude equation (\eqref{vamaglin}) is an equality constraint.

The remaining three constraints are OLTC tap position constraints. 
$\tau_{p,t}\in\mathbb{Z}$ indicates the tap position is an integer, where $\tau_{p,t}$ denotes the tap position of OLTC $p$ at time step $t$ and $\mathbb{Z}$ represents integer numbers. $\tau_{p,\min}\le\tau_{p,t}\le\tau_{p,\max}$ denotes the tap position is within [$\tau_{p,\min}$  $\tau_{p,\max}$], $\tau_{p,\min} = -16$ and $\tau_{p,\max} = 16$ are the minimum and maximum tap positions of OLTC $p$, respectively. $|\tau_{p,t}-\tau_{p,t-1}|\le \Delta \rm TO_{p,\max}$ constrains tap position change within two consecutive time steps. Since OLTCs react slowly, $\Delta \rm TO_{p,\max}$ (1 tap operation per 30~sec) avoids unrealistic tap operation changes by limiting the maximum tap operations between two consecutive time steps.

\section{Case Study}
\label{sec:casestudy}

\subsection{Distribution Feeder Models}

\begin{table}[]
\normalsize
\centering
\caption{Test Feeder Properties.}
\label{feeder}
\begin{tabular}{l|c|c}
\hline
Feeder              & \multicolumn{1}{l|}{IEEE 37} & \multicolumn{1}{l}{Utility Feeder} \\ \hline
\# of nodes         & 120                              & 2844                                   \\
\# of Loads         & 30                               & 584                                    \\
\# of PV            & 30                               & 203                                    \\
Peak Load (MVA)     & 2.73                             & 8.50                                   \\
PV Penetration (\%) & 150                              & 150                                    \\
\# of OLTC          & 1                                & 1                                     
\end{tabular}
\end{table}

\begin{figure} 
\centering
\includegraphics[width=0.44\textwidth]{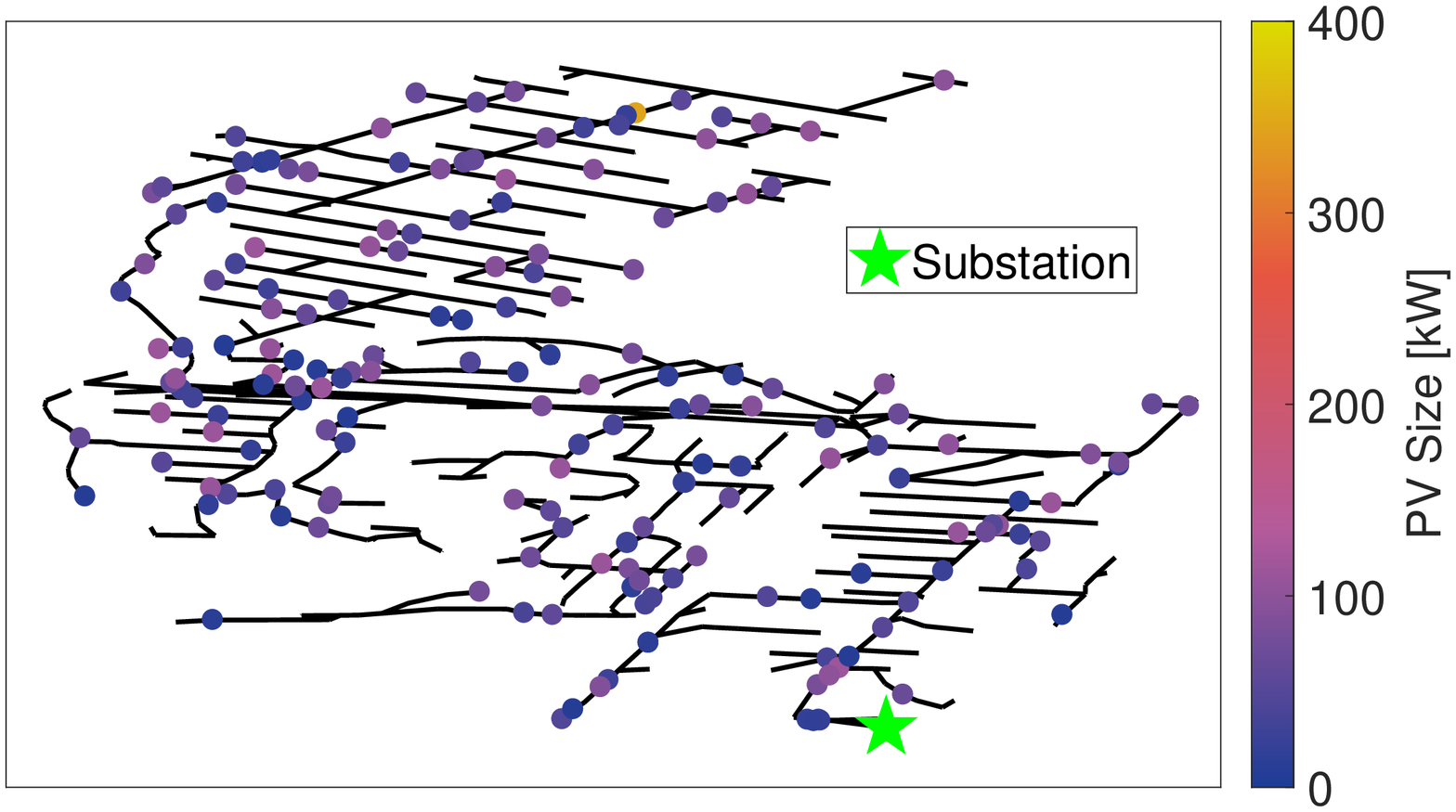}
\caption[]{Feeder topologies of the California utility feeder. Black lines represent feeder lines. Each dot is a PV system and its color indicates its AC power rating. The OLTC is located at the substation (green star).}
\label{pointloma}
\end{figure}

To evaluate the proposed method, quasi-steady state simulations are carried out on two multi-phase unbalanced feeders, the modified IEEE 37 bus feeder and a real California utility feeder (feeder 10 in \cite{li2016distribution} as shown in Fig.~\ref{pointloma}). The feeder properties are summarized in Table~\ref{feeder}. The IEEE 37 bus feeder is simulated for 24 hours with 30-sec resolution using measured solar profiles from a partially cloudy day. Due to computation time limitations for the large utility feeder, a 5-min simulation time step is used. The simulation is performed for a clear day.

For the IEEE 37 bus feeder, 30 PVs with DC power rating ranging from 23 to 206~kW and totalling $P_{\rm pv}^{\rm peak}=4.1$~MW  are randomly deployed on the feeder. The total PV penetration on the feeder is 150\% by capacity:
$PV_{\rm Pen}=\frac{P_{\rm pv\_peak}}{P_{\rm load\_peak}} \times 100\%$.
340 PVs with DC ratings varying from 7~kW to 458~kW are connected to the utility feeder, resulting in a total capacity $P_{\rm pv\_peak} = 12.8$ MW and also 150\% PV penetration. 10\% oversizing of AC power rating is assumed for the SIs on both feeders \cite{turitsyn2011options}. 
Both feeders contain one OLTC at the substation. The OLTC tap position can vary from $\tau=-16$ to +16 with voltage regulation capability of [0.9 1.1] p.u..

\subsection{Voltage Regulation Methods}
\label{VoltReg}
The proposed method is benchmarked against the widely-used conventional autonomous voltage regulation scheme (AVR). The two different voltage regulation strategies are summarized in Table~\ref{VoltRegMethod}.

\subsubsection{Autonomous Voltage Regulation (AVR)} 

\begin{figure} 
\centering
\includegraphics[width=0.40\textwidth]{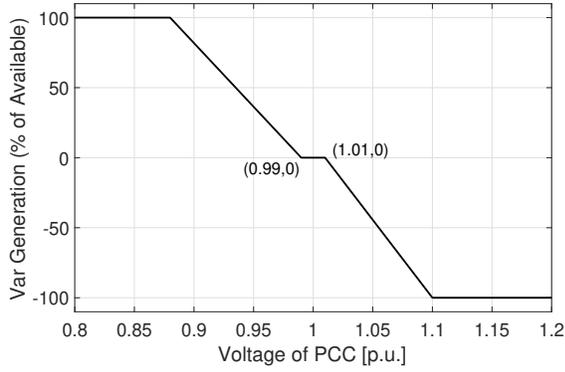}
\caption[]{Volt-Var curve of SI adapted from \cite{california2014recommendations}. The SI injects/absorbs the corresponding percentage of available vars based on the voltage at the point of common coupling (PCC). The available vars ($Q_{max}$) are limited by PV real power generation and SI rating as shown in Fig.~\ref{fig:SI}.}
\label{VoltVar}
\end{figure}

 In AVR, voltage control devices operate autonomously based on pre-defined rules/curves without coordination with each other. SIs output reactive power following the Volt-Var curve in Fig.~\ref{VoltVar}, which is the recommended default Volt-Var curve by the California Public Utility Commission \cite{california2014recommendations}. OLTCs change tap position to keep the deviation of the local busbar voltage from the preset reference voltage within certain limits. The OLTC reference voltage is set to 1.03~p.u. for IEEE 37 bus feeder and 1.02~p.u. for the utility feeder. The voltage regulation bandwidth is 0.0167~p.u. for both feeders. For better voltage regulation, the tap time delay is set to be 0 sec. All other OLTC parameters are set as the default OpenDSS \cite{OpenDSS} values.

\subsubsection{Optimal Voltage Regulation (OVR)}

\label{OTC}
For OVR, OLTCs and SIs are coordinated to minimize voltage deviations as described in Section~\ref{sec:opt}. Per the optimization outputs, SIs participate in voltage regulation via reactive power absorption and injection and OLTC tap position are specified. A reference voltage is not needed as the OLTCs will follow the optimal tap position. 

\begin{table}[]
\normalsize
\centering
\caption{Summary of autonomous voltage regulation (AVR) and optimal voltage regulation (OVR).}
\label{VoltRegMethod}
\begin{tabular}{c|c|c}
\hline
                & OLTC                 & PV            \\ \hline
AVR (benchmark) & Autonomous control & Volt-var Curve      \\ \hline
OVR (proposed)  & Tap optimized      & VAR optimized \\ \hline
\end{tabular}
\end{table}

\subsection{Selection of Weighting Factors for OVR}
\label{sec:weightfactor}

Since the optimization objective $J$ is a weighted sum of $J_1$ and $J_2$, heavier weighting on $J_1$ will improve the voltage profile at the cost of more TOs and vice versa. Therefore, appropriate weighting factors should be chosen to achieve a desired trade-off between voltage profile and number of TOs. Several combinations of weighting factors ($w_1$, $w_2$) are tested with simulations on the IEEE 37 bus feeder on 150\% PV penetration (Table \ref{weightfactors}). 

As expected, larger weighting factors ($w_2$) on $J_2$  cause decreasing total TO, while voltage deviations generally increase. Relative to $w_2 <= 0.05$, $w_2$= 0.15 provides a large reduction in TO without a significant increase in voltage deviations. And simulations with $w_2$ = 0.15 on the California utility feeder also show minimization of voltage deviations with a reasonable number of total TOs. Therefore, $w_2$ = 0.15 is used for both test feeders hereinafter.

The best combination of $w_1$ and $w_2$ for different feeders may vary due to the operator preferences between better voltage regulation or less TO, different locations of OLTCs, feeder topologies, distribution of PVs, etc. Local adjustments of the weighting factors are therefore recommended.

\begin{table}[]
\normalsize
\centering
\caption{Case study of different objective function weights $w_2$ on the IEEE 37 bus feeder with 150\% PV penetration. $w_1$ is fixed at 1. The mean voltage deviation is calculated for 8:00 - 17:00 of all nodes. The total TO is counted for the 24 hour day.}
\label{weightfactors}

\begin{tabular}{c|ccc}
\hline
$w_1$                & $w_2$      & Mean Voltage Deviation (p.u.) & Total TO \\ \hline
\multirow{6}{*}{1} & 0.001 & 0.0170          & 94      \\ \cline{2-4}                  
                   & 0.010   & 0.0170          & 91      \\ \cline{2-4} 
                   & 0.050    & 0.0172          & 38      \\ \cline{2-4} 
                   & 0.150    & 0.0174          & 2       \\  \hline
\end{tabular}
\end{table}

\section{Distribution Feeder Simulations Results}
\label{sec:result}
\subsection{Voltage Estimation Accuracy}
\label{acc}
\begin{figure} 
\centering
\includegraphics[width=0.42\textwidth]{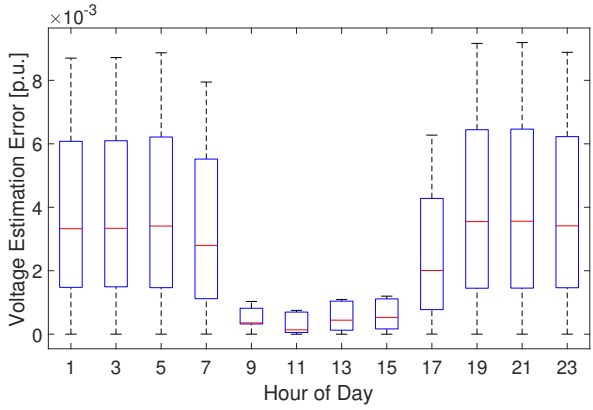}
\caption[]{Box plot of voltage estimation errors for the IEEE 37 bus test feeder. For readability, the results are aggregated over two hours into 12 groups.}
\label{volterr}
\end{figure}

Given that estimated node voltages are used in the optimization to determine optimal OLTC tap position and SI reactive power, we examine the errors resulting from the linearization of feeder nodal voltage equations (\eqref{eq:VI_derivaive}), the admittance matrix (Section~\ref{subsec:VoltEffectTO}), power injection constraints (Section~\ref{subsec:VoltEffectCurr}), and the voltage magnitude (\eqref{vamaglin}). Errors are defined as the differences in estimation of voltage magnitude from \eqref{vamaglin} versus the non-linear AC power flow results from OpenDSS (Fig.~\ref{flowchart}) 
\begin{align}E(t)_i = V_{\rm estimate}(t)_i - V_{\rm OpenDSS}(t)_i.
\end{align}
Fig.~\ref{volterr} presents $E(t)$ distributions for the IEEE 37 bus feeder. Since the voltage estimations match the AC power flow results closely, it can be concluded that the proposed model estimates voltage magnitudes accurately. The maximum error magnitude is 0.009~p.u. and the mean absolute error magnitude is always under 0.004~p.u.. Around noon when the distribution feeder is prone to voltage violations with high PV generation, the voltage estimation errors are less than 0.001~p.u.. At night when the estimation errors are relatively high, the distribution system is much less likely to experience voltage violations even with simple local Volt-Var functions (Fig.~\ref{fig:voltdevdist}). And the proposed method can improve the night time voltage profile further as shown in Fig.~\ref{fig:voltdevdist}. Therefore, the proposed method can effectively mitigate voltage violations despite somewhat larger voltage estimation errors at night.

\subsection{Performance Under Perfect Forecasts}
\subsubsection{IEEE 37 Bus Feeder}
\label{subsec:res:ieee37}
\subsubsection*{Voltage Profile}
\begin{figure}
\centering
   \begin{subfigure}[b]{0.46\textwidth}
  \includegraphics[width=0.99\textwidth]{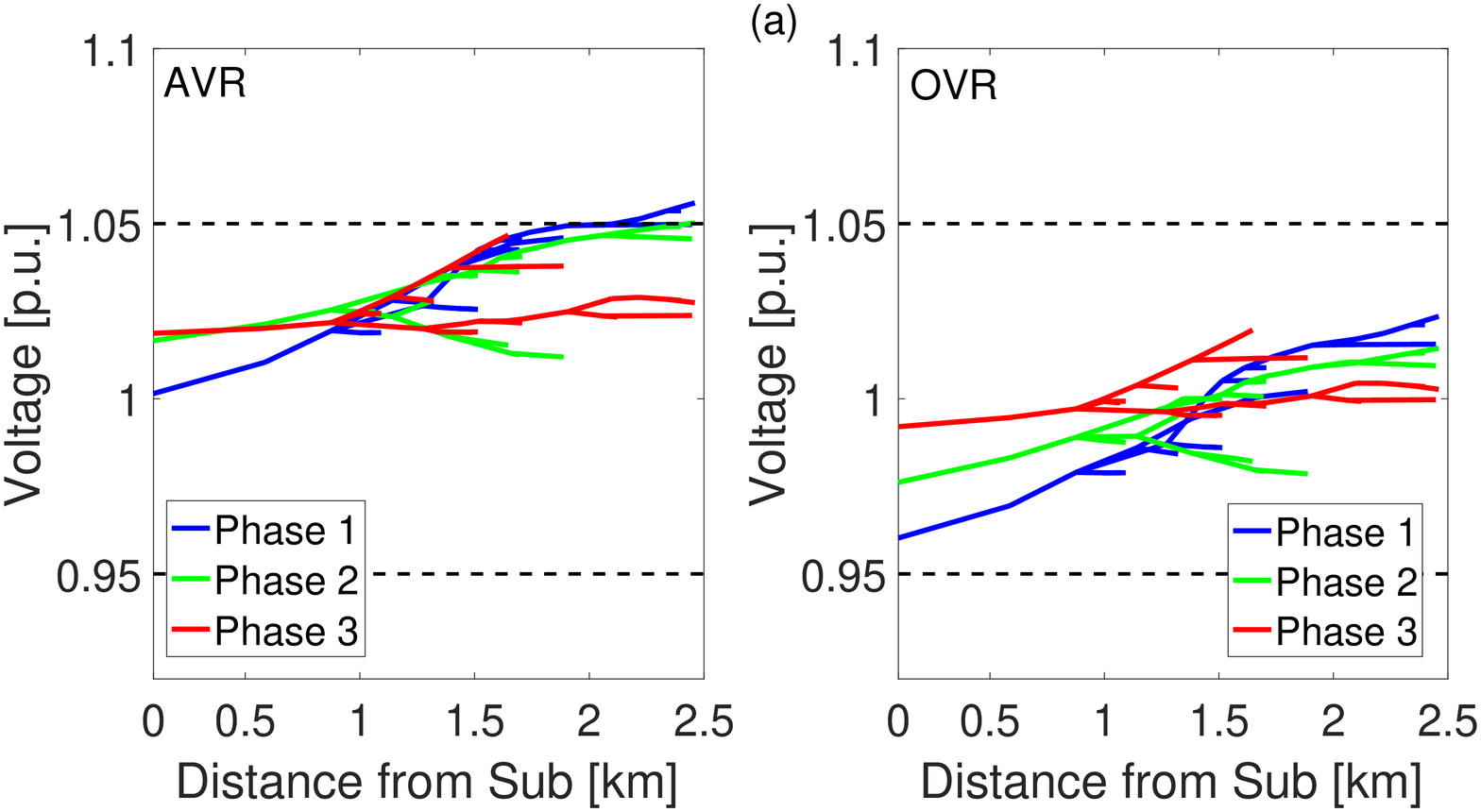}
   \label{Volt12} 
\end{subfigure}

\begin{subfigure}[b]{0.46\textwidth}
\includegraphics[width=0.99\textwidth]{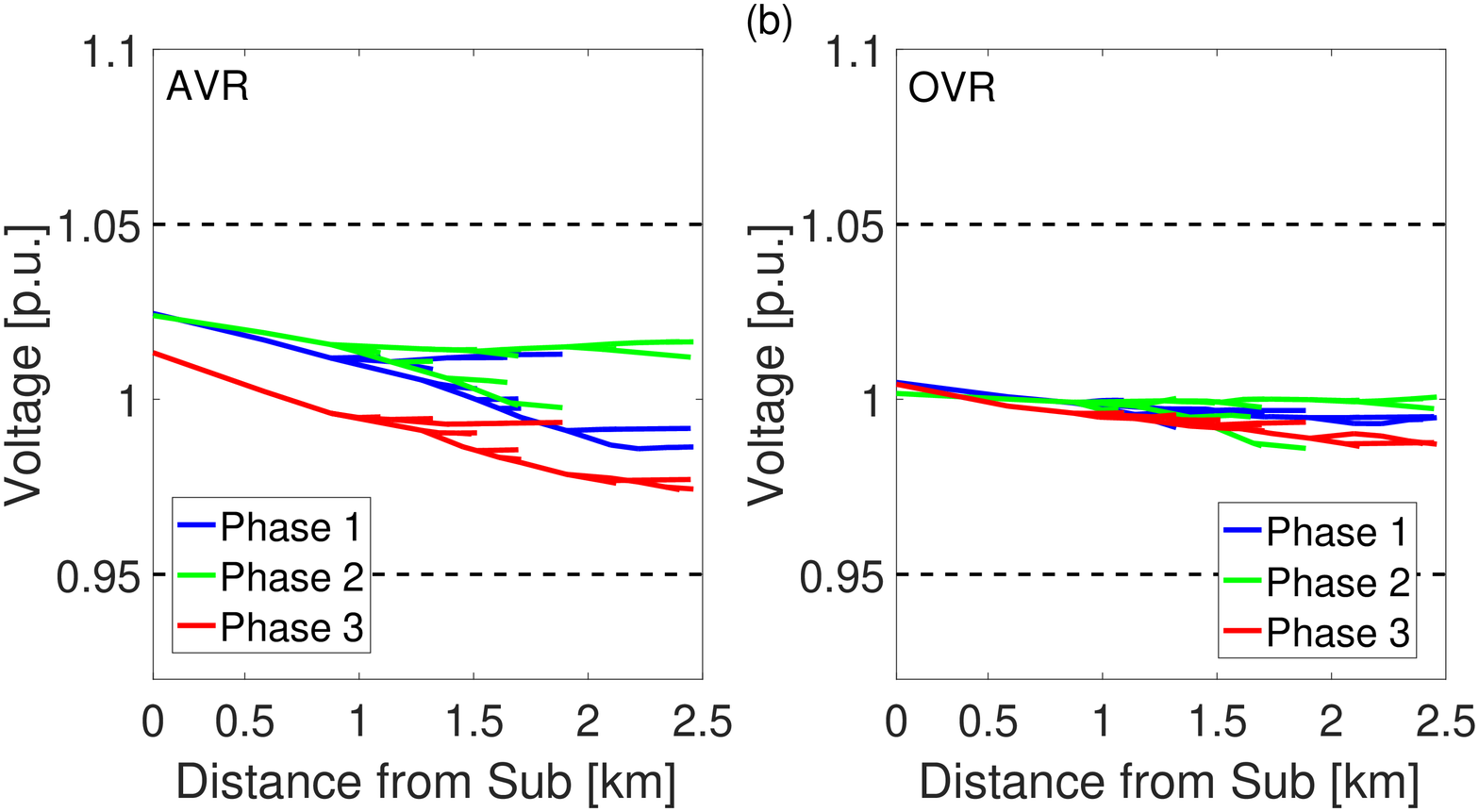}
   \label{Volt21}
\end{subfigure}

\caption[Voltage along the feeder]{Feeder voltage profile at 11:32 (top, a) and 21:00 (bottom, b) for the AVR (left) and OVR (right) voltage regulation methods.
\vspace{-2ex}
\vspace{-0ex}}
\label{voltalongfeeder}
\end{figure}

Fig.~\ref{voltalongfeeder} presents snapshot voltage profiles of the feeder around noon (11:32~h, medium loading of 0.84~MVA, large PV generation of 3.3~MW) and in the evening (21:00~h, heavy loading of 1.91 MVA, no PV generation). At noon, a voltage increase along the feeder results from reverse power flow caused by excess PV production. For AVR, the voltage violates ANSI standards at the feeder end at 1.056~p.u. \cite{std2011c84}. The over-voltage violation occurs in phases 1 and 2, while the voltage on phase 3 remains within [0.95 1.05] p.u.. At the feeder end, there are also large voltage imbalances across different phases with a mean voltage imbalance of 0.012~p.u. and a max of 0.030~p.u.

On the contrary, OVR keeps the voltage of all phases within the [0.95 1.05]~p.u. ANSI limits. Due to limited available reactive power capacity around noon (at full active power, $Q_{\rm max}=42\%$ of the rated power), the mean voltage unbalance stays the same and the max unbalance increase to 0.041~p.u. as a result of correcting the over-voltage issues. If PV curtailment was allowed, more reactive power support would reduce the voltage unbalance since OVR minimizes total voltage deviation, bringing all the voltages closer to 1~p.u.. 

At 21:00 heavy loading causes a large voltage drop with AVR. Again, voltage discrepancies between phases are large: the largest voltage difference occurs between phases 1 and 3 at the feeder end at 0.042~p.u., equivalent to 42\% of the allowable voltage range. The mean voltage unbalance is 0.017 p.u.. With OVR, the voltages remain close to 1~p.u. across the entire feeder, resulting in a more desirable homogeneous (flat) voltage profile. The voltage unbalance is substantially reduced with a maximum of 0.014~p.u. (a 67\% reduction compared to AVR) and a mean of 0.005 p.u.. OVR squeezes the voltage range on all phases toward 1 p.u. with coordinated reactive power support from SIs, reducing the voltage unbalance on the feeder. The favorable OVR results are enabled by coordination of OLTCs and SIs through optimization and also by unlimited reactive power support (Fig.~\ref{fig:SI}) at night.

\begin{figure}
\centering
   \begin{subfigure}[b]{0.46\textwidth}
  \includegraphics[width=0.99\textwidth,trim={0cm 0cm 0cm 0cm}]{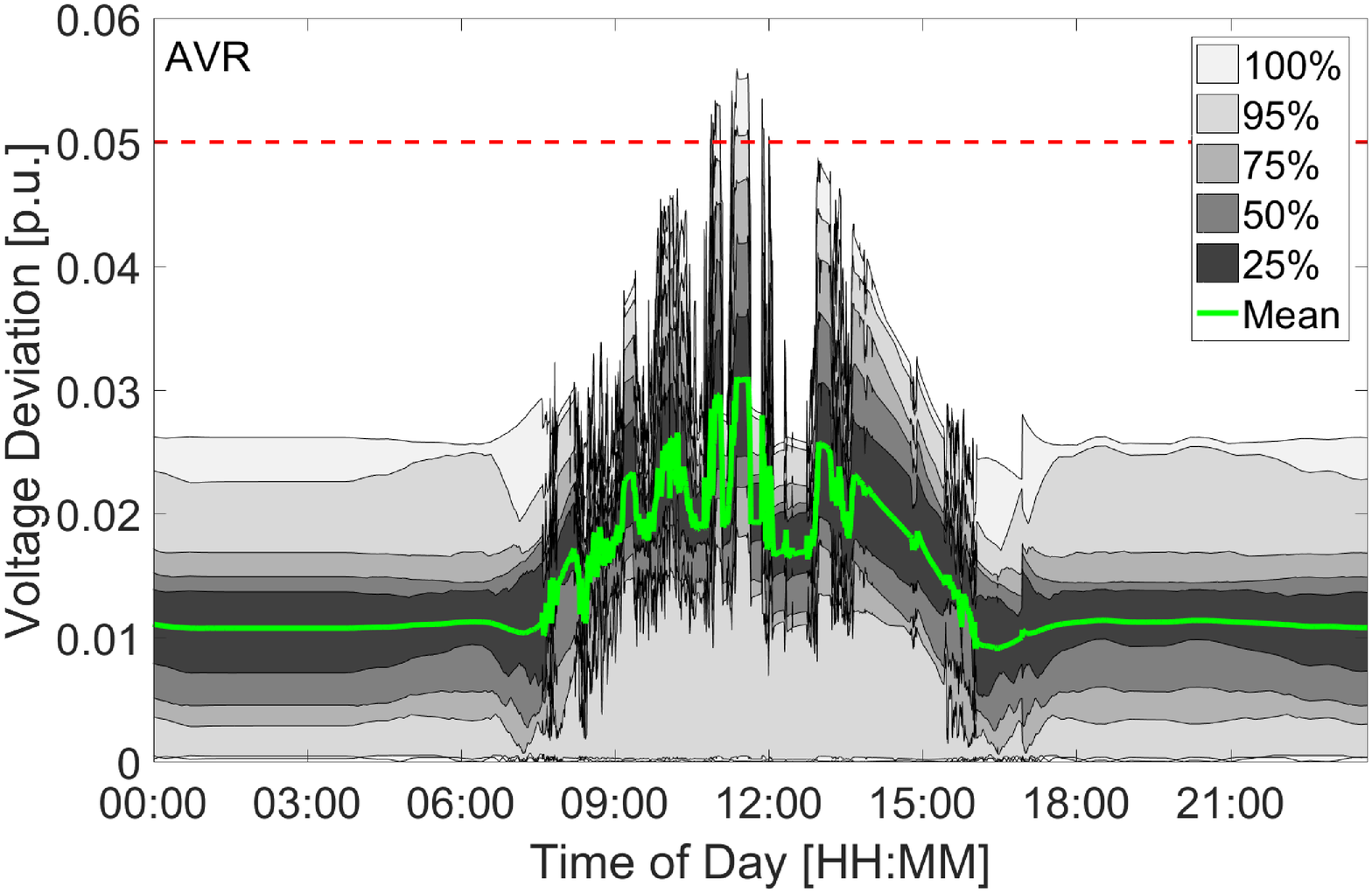}
\end{subfigure}

\begin{subfigure}[b]{0.46\textwidth}
\includegraphics[width=0.99\textwidth,trim={0cm 0cm 0cm 0.1cm}]{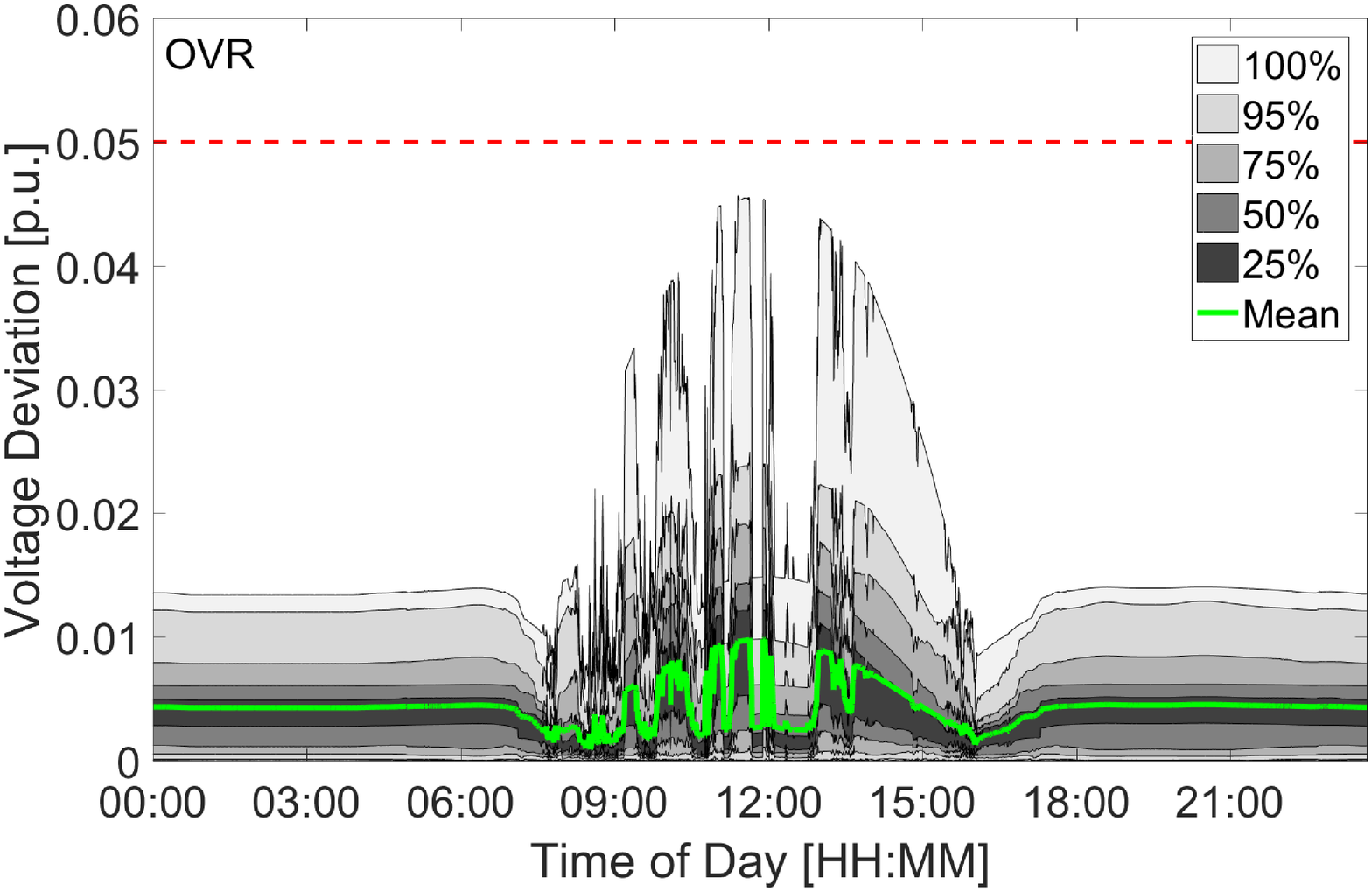}
\end{subfigure}

\caption[Voltage along the feeder]{Distribution of the absolute values of feeder nodal voltage deviations from $1$ p.u. for all nodes on the IEEE 37 bus feeder. top: AVR; bottom: OVR. The absolute values are plotted for consistency with $J_1 \eqref{J1}$. The red dashed line represents the [0.95 1.05]~p.u. ANSI voltage limits. For the IEEE 37 bus feeder, all voltage deviations greater than 0.05 are over-voltages.
\vspace{-2ex}
\vspace{-0ex}}
\label{fig:voltdevdist}
\end{figure}

Fig. \ref{fig:voltdevdist} compares absolute voltage deviation of all nodes on the feeder between AVR and OVR. For AVR, the mean voltage deviation is around 0.011~p.u. during periods without PV production. The voltage deviation mean increases when PV power production ramps up starting around 08:00 and reaches 0.031 p.u. at noon. Generally, voltages are scattered far around the mean value and the range is larger near noon due to high PV generation. The voltage deviations exceeded the 0.05 p.u. limit (over-voltage) for over one hour near noon. With OVR, the voltage deviations of all nodes are under the 0.05 p.u. limit, eliminating the over-voltage problems in the AVR case. The voltage deviation mean decreases to below 0.005~p.u. at night and is always below 0.01 p.u. during the day. The minimum average voltage deviations around 08:30 and 16:00 occur when PV generation balances load consumption minimizing power flow on the feeder. Like for AVR, the OVR voltage deviation is larger during day time and peaks around noon. The noon peak is a combined effect of maximum power flow on the feeder (i.e. larger voltage change) caused by more PV power production and reduced available reactive power of SIs (Fig.~\ref{fig:SI}).\\

\subsubsection*{OLTC Tap Operations}
\begin{figure} 
\centering
\includegraphics[width=0.42\textwidth]{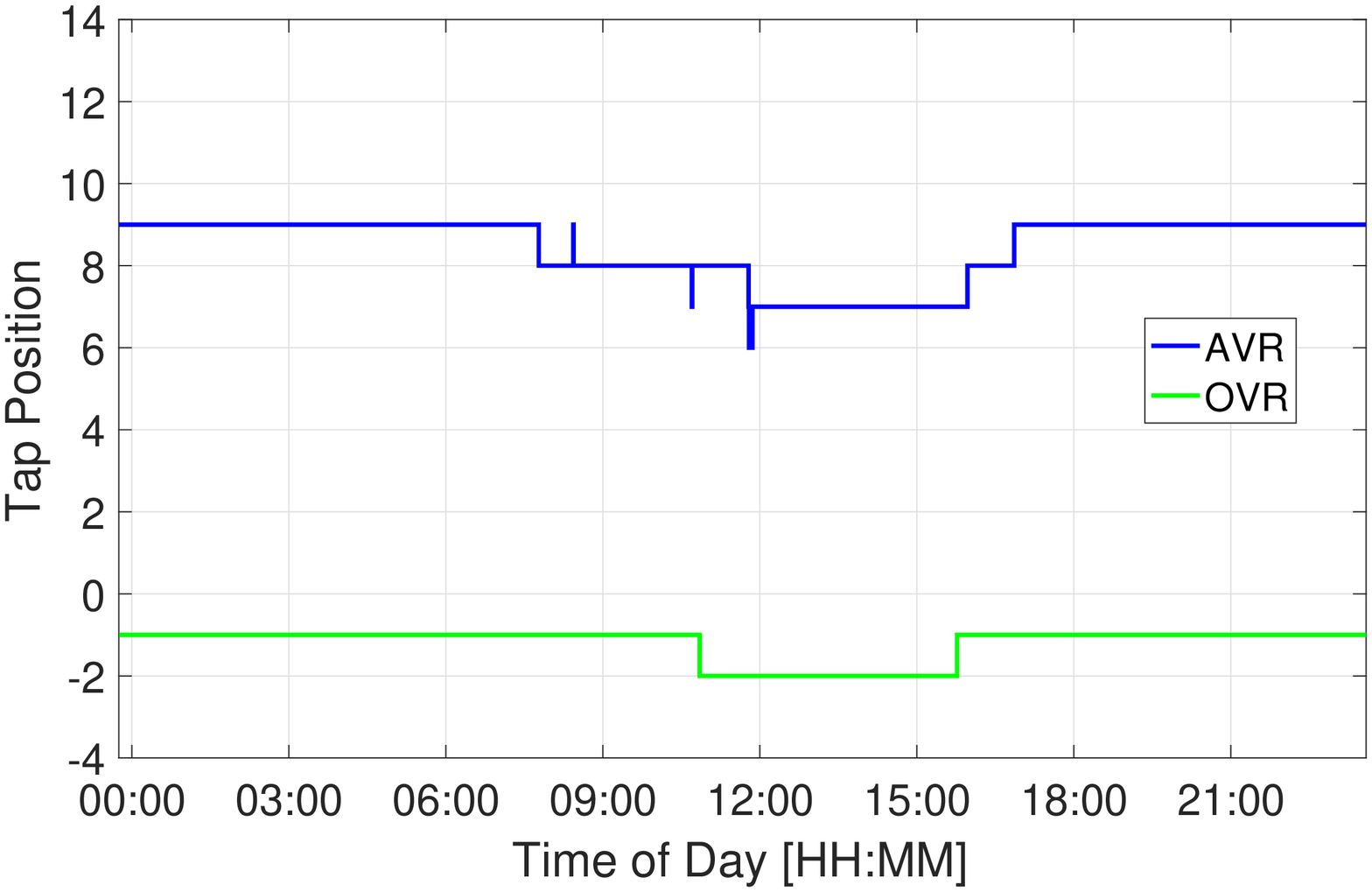}
\caption[]{Time series of OLTC tap positions $\tau$ for the IEEE 37 bus feeder.}
\label{oltc}
\end{figure}

Fig.~\ref{oltc} presents the OLTC tap positions. For AVR, the tap position is set high (+9) during the night with no PV production compensating the voltage drop on the feeder. During the PV production period, the tap position is lowered (+6) due to increasing voltage. Since the OLTC and SIs operate autonomously based on local voltage without coordination, the fluctuating PV generation (as indicated by Fig.~\ref{fig:voltdevdist}) triggers three immediate up-and-down tap operations. In total, 10 tap operations occur during the day. With OVR, SIs provide optimized reactive power support in coordination with OLTCs, allowing a lower tap position setting without violating operation limits. Unnecessary tap operations are also avoided per objective $J_2$ (\ref{J2}) resulting in only 2 total tap operations, which is an 80\% reduction when compared to AVR.

\subsubsection{Large California Utility Feeder}
\begin{figure}
\centering
   \begin{subfigure}[b]{0.41\textwidth}
  \includegraphics[width=0.99\textwidth,trim={0cm 0cm 0cm -0.05cm}]{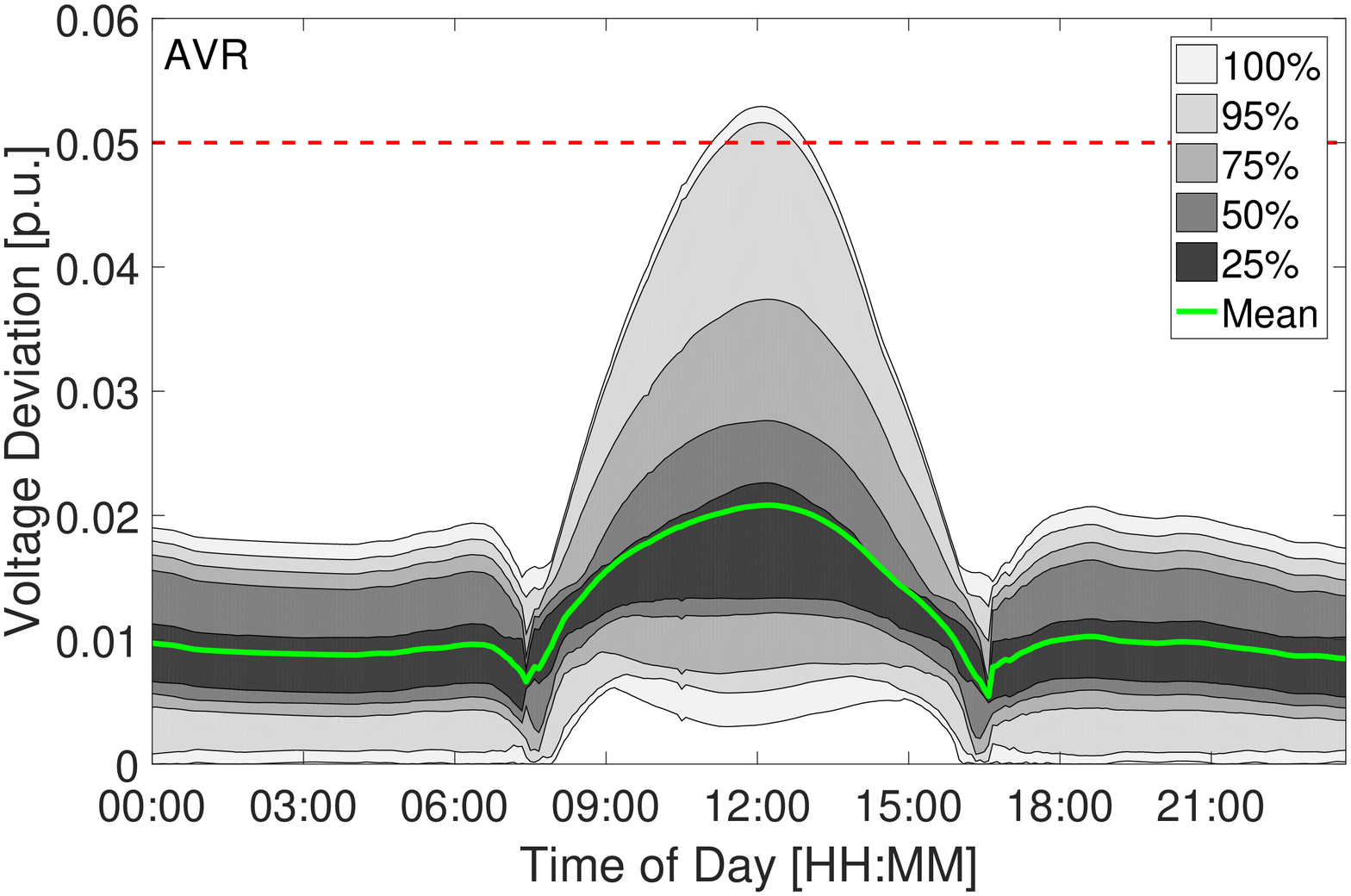}
   
\end{subfigure}

\begin{subfigure}[b]{0.41\textwidth}
\includegraphics[width=0.99\textwidth,trim={0cm 0cm 0cm 0.1cm}]{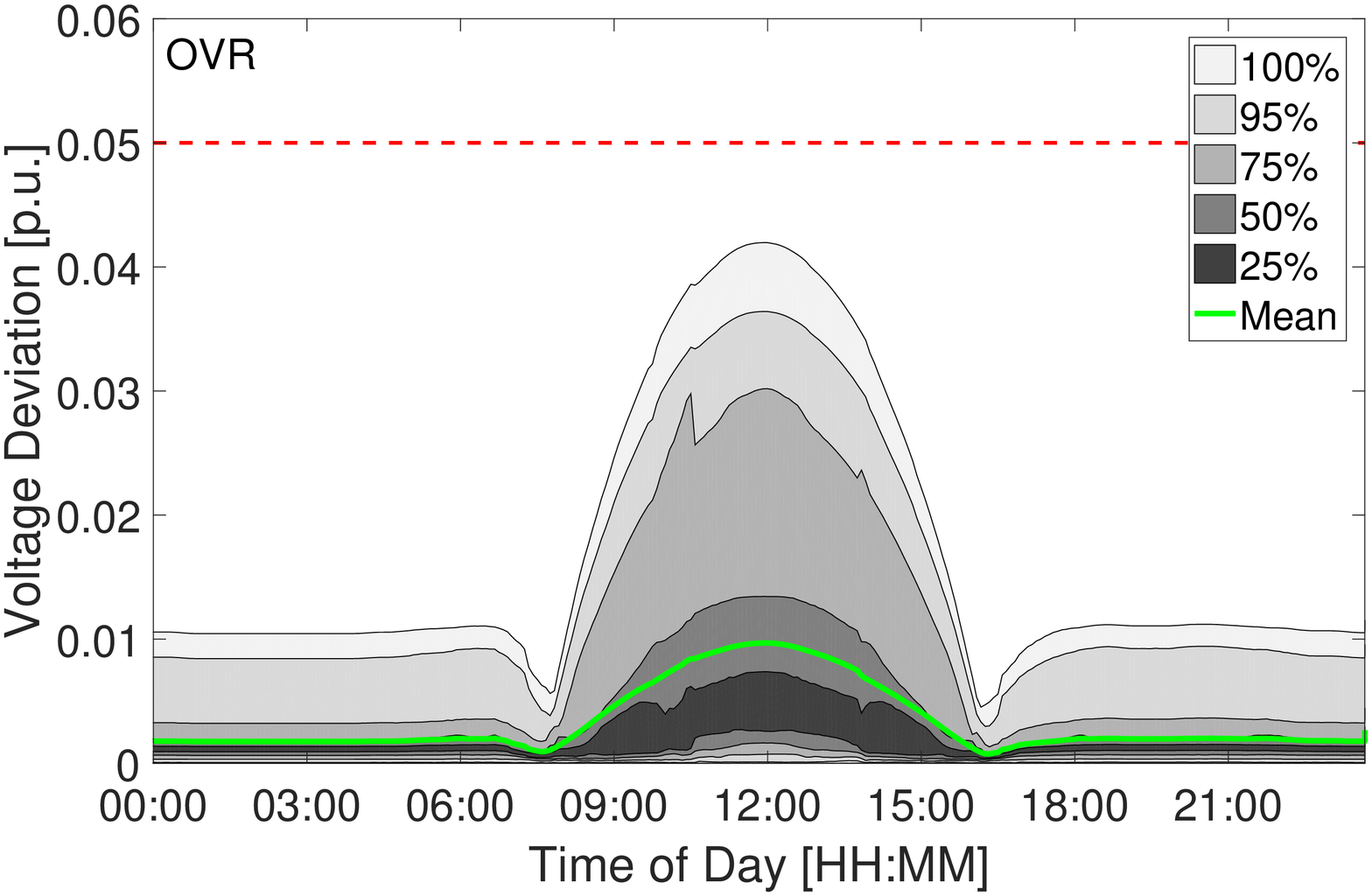}
\end{subfigure}

\caption[Voltage along the feeder]{Distribution of absolute values of feeder nodal voltage deviation from $1$ p.u. for AVR (top) and OVR (bottom) cases. 
\vspace{-2ex}
\vspace{-0ex}}
\label{voltdevdist_pointloma}
\end{figure}

Fig.~\ref{voltdevdist_pointloma} displays the voltage profile of the large utility test feeder with AVR and OVR. The mean AVR voltage deviation is around 0.010~p.u. at night and increases to 0.021 p.u. around noon. The feeder experiences over-voltages for about 2 hours when PV generation peaks. The largest voltage deviation is 0.053~p.u., corresponding to a 1.053 p.u. overvoltage.
With OVR, the average nodal voltage deviation reduces to 0.002~p.u. at night and reduces to 0.010~p.u. near noon. All voltage deviations are under the 0.05~p.u. limit, indicating that the over-voltage issues for AVR are eliminated. The maximum voltage deviation decreases to 0.042 p.u., corresponding to a maximum voltage of 1.042 p.u.. Due to smooth PV generation of clear day the OLTC did not operate under AVR despite over-voltage issues on the feeder. With coordinated OLTC and SIs under OVR, 2 tap operations suffice to resolve the over-voltage problem for the entire day.

\subsection{Performance With Forecast Errors}
\label{subsec:resforecast}

\begin{figure} 
\centering
\includegraphics[width=0.44\textwidth]{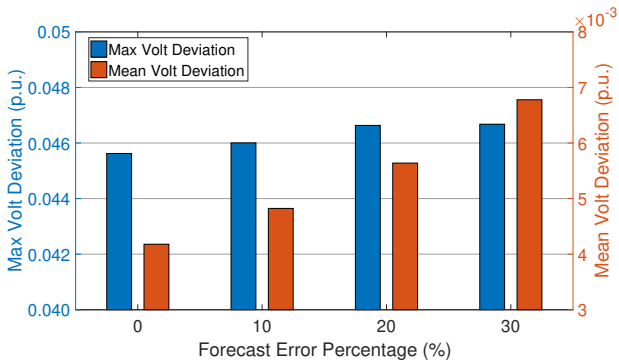}
\caption[]{Maximum and average voltage deviation of the IEEE 37 bus feeder under different levels of forecast errors with OVR. Both quantities are calculated based on all nodal voltages for simulations of the same cloudy day as Section \ref{subsec:res:ieee37}}
\label{fig:voltdev_errtest}
\end{figure}

To evaluate the robustnes of the proposed OVR, multiple case studies are carried out on the IEEE 37 bus feeder assuming different levels of forecasting error. The PV and load profiles with forecast errors are constructed using the following equation:
\begin{align}
    y(t) = (1 + \alpha\epsilon(t))y_{true}(t),
    \label{eq:forecasterr}
\end{align}
where $y(t)$ represents the forecasted PV output or load consumption at time step $t$, $y_{true}(t)$ is true PV/load profile for time step $t$. $\alpha$ is a scalar to represent the maximum error ratio in terms of $y_{true}(t)$ while $\epsilon(t)$ is a random number from an uniform distribution within [-1,1].

Fig.~\ref{fig:voltdev_errtest} presents the maximum and average nodal voltage deviations within 24 hr simulation period for IEEE 37 bus feeder with OVR for $\alpha$ = [0, 0.1, 0.2, 0.3]. $\alpha$ = 0 indicates perfect forecasts as in Section \ref{subsec:res:ieee37}.

As shown in Fig.~\ref{fig:voltdev_errtest}, the both maximum and average voltage deviations increase with larger forecasting errors. This is consistent with expectations that less accurate forecasts will lead to performance deterioration. Even with 30\% forecast error, the OVR is still capable of keep the maximum voltage deviation (0.0467 p.u.) under 0.05 p.u., i.e. without voltage violations. In contrast, AVR fails to maintain e the voltage within [0.95 1.05] p.u. operation limits as shown in Fig.~\ref{fig:voltdevdist} (Note that since AVR does not require any forecasts, it is always tested with true PV and load profiles). With AVR, the maximum voltage deviation is 0.056 p.u., corresponding to an over-voltage of 1.056 p.u.. As for mean nodal voltage deviation, OVR keeps average voltage deviation under 0.0068 p.u. even with a forecast error of 30\% while the average voltage deviation is 0.0135 p.u. for the AVR case. OVR significantly decreases voltage deviation per $J_1$.
\\
\subsection{Computation Time}
Solving the optimization problem at an operational timescale would enable the control to be used in real time applications. 
Table~\ref{ct} compares the average computation time in \cite{othman2019coordinated,guggilam2016scalable} for the IEEE 33 bus feeder and IEEE 2500 node feeder with the OVR cases in this paper. For the IEEE 33 bus feeder with 99 nodes, 3 DGs, 2 ShCs and 1 OLTC, the average computation time is 40~s in \cite{othman2019coordinated} using an Intel Core i7-2600 @ 3.4 GHz processor. With the proposed OVR algorithm in this paper, however, the solution time is less than 1 second for the slightly larger IEEE 37 feeder with 120 nodes, 30 PVs and 1 OLTC using an Intel(R) Core(TM) i7-4700MQ 2.8-GHz processor. For the large feeders, the solution time for IEEE 2500 node feeder in \cite{guggilam2016scalable}
is 600~s with an Intel core i7-4710HQ @ 2.5 GHz processor, while our proposed OVR reduces computing time by 80\% for a larger feeder on the PC with Intel (R) Core(TM) i7-4700MQ 2.8-GHz processor. While the computation cost comparison is not apples-to-apples, the results strongly favor our OVR approach which outperforms prior research by a large margin on larger feeders and comparable and even inferior computing resources.

Due to compatibility issue between CVX \cite{cvx} and the newest version of Gurobi solver \cite{gurobi}, Gurobi v6.5 is used in this paper; the solution speed could be further improved with the latest version Gurobi v8.1  \cite{gurobibench}.

\begin{table}[]
\normalsize
\centering
\caption{Comparison of average computation time (s) per time step in \cite{othman2019coordinated,guggilam2016scalable} and OVR case study in this paper.}
\label{ct}
\begin{tabular}{c|c|c|c}
\hline
Test Feeder    & \# of nodes & Optimization & Solution time (s) \\ \hline
IEEE 33        & 99          & {[}15{]}     & 40                \\ \hline
IEEE 37        & 120         & OVR          & 0.95              \\ \hline
IEEE 2500      & 2500        & {[}9{]}      & 600               \\ \hline
Utility Feeder & 2844        & OVR          & 127               \\ \hline
\end{tabular}
\end{table}

\section{conclusions}
\label{sec:conc}

A novel method of coordinating OLTCs and SI reactive power for voltage regulation was proposed. OVR is capable of coordination voltage regulation between multiple OLTCs and SIs. The proposed OVR is compared against conventional AVR through simulations on the highly unbalanced IEEE 37 bus test  and a large California utility feeder. 
Results show that OVR can mitigate over-voltage violations, significantly reduce voltage deviations, decrease voltage unbalance across phases, and avoid unnecessary tap operations. This is achieved by effective coordination between OLTCs and SI reactive power control. The robustness of the proposed OVR is also demonstrated on the IEEE 37 bus test feeder assuming different levels of forecast error. The computational efficiency of OVR is superior to prior methods and the solution time is compatible with real-time operation on large California utility feeder with 2844 nodes and 203 PVs even on a regular PC. 

\bibliographystyle{IEEEtran}
\bibliography{Main.bib}

\end{document}